\documentclass[12pt]{article}
\usepackage{jhep-mod}
\usepackage{bm}
\usepackage{amssymb,amsmath,amsthm}
\usepackage{mathrsfs}
\usepackage[utf8]{inputenc}
\usepackage{enumerate}
\hypersetup{colorlinks=true} 
\usepackage{xcolor}
\usepackage{appendix}
\usepackage{graphicx}
\usepackage{dcolumn}
\usepackage{bm}
\usepackage{multirow}
\usepackage{float}
\usepackage{tikz}
\usepackage{colortbl}
\usepackage[normalem]{ulem}
\definecolor{purple}{rgb}{1,0,1}
\definecolor{lime}{HTML}{A6CE39} 

\newcommand{\orcidicon}{%
	\begin{tikzpicture}
	\draw[lime, fill=lime] (0,0) 
		circle [radius=0.16] 
		node[white] {{\fontfamily{qag}\selectfont \tiny ID}};
	\draw[white, fill=white] (-0.0625,0.095) 
		circle [radius=0.007];
	\end{tikzpicture}
	\hspace{-2mm}
}
\newcommand\orcidMatt{{\href{https://orcid.org/0000-0003-1088-6485}{\orcidicon}}}
\newcommand\orcidAlex{{\href{https://orcid.org/0000-0002-1763-3563}{\orcidicon}}}

\begin{document}

\title{\null\vspace{-85pt} \huge ISCOs and OSCOs in the presence of positive cosmological constant}

\author{
\Large Petarpa Boonserm$\,^{1,3}$, Tritos Ngampitipan$\,^{2,3}$, \\
Alex Simpson$\,^{4}$\orcidAlex\!\!, {\sf  and} Matt Visser$\,^{4}$\orcidMatt}
\affiliation{
$^{1}$  Department of Mathematics and Computer Science, Faculty of Science,\\
\null\qquad Chulalongkorn University,  Bangkok 10330, Thailand.  
}
\affiliation{
$^{2}$  Faculty of Science, Chandrakasem Rajabhat University,
 Bangkok 10900, Thailand.
}
\affiliation{$^{3}$ Thailand Center of Excellence in Physics,
Ministry of Higher Education, Science, \\
\null\qquad
Research, and Innovation, 328 Si Ayutthaya Road,  Bangkok 10400, Thailand
}
\affiliation{$^{4}$ School of Mathematics and Statistics, Victoria University of Wellington, \\
\null\qquad PO Box 600, Wellington 6140, New Zealand}
\emailAdd{petarpa.boonserm@gmail.com}
\emailAdd{tritos.ngampitipan@gmail.com}
\emailAdd{alex.simpson@sms.vuw.ac.nz}
\emailAdd{matt.visser@sms.vuw.ac.nz}

\abstract{
\parindent0pt
\parskip7pt
Normally one thinks of the observed cosmological constant as being so small that it can be utterly neglected on typical astrophysical scales, only affecting extremely large-scale cosmology at Gigaparsec scales. Indeed, in those situations where the cosmological constant  only has a  \emph{quantitative} influence on the physics, a separation of scales argument guarantees the effect is indeed negligible. The exception to this argument arises when the presence of a cosmological constant \emph{qualitatively} changes the physics. One example of this phenomenon is the existence of \emph{outermost stable circular orbits} (OSCOs) in the presence of a positive cosmological constant. 
Remarkably the size of these OSCOs are of a magnitude to be astrophysically interesting. For instance: for galactic masses the OSCOs are of order the inter-galactic spacing, for galaxy cluster masses the OSCOs are of order the size of the cluster. 

\smallskip
{\sc Date:} 15 Sept 2019; 23 Sept 2019; 30 Jan 2020; \LaTeX-ed \today

innermost conceivable circular orbit (ICCO); outermost conceivable circular orbit (OCCO);\\
cosmological constant. 

\smallskip
%
{\sc arXiv:} 1909.06755 \\
{\sc published as:} Physical Review D {\bf 101 \# 2} (2020) 024050\\
{\sc published title:}
Innermost and outermost stable circular orbits in the presence of a positive cosmological constant.
\\
{\sc url:} \url{https://link.aps.org/doi/10.1103/PhysRevD.101.024050}\\
{\sc doi:} 10.1103/PhysRevD.101.024050
}

\maketitle


\def\tr{{\mathrm{tr}}}
\def\diag{{\mathrm{diag}}}
\parindent0pt
\parskip7pt

\def\rocco{r_\mathrm{\scriptscriptstyle OCCO}}
\def\ricco{r_\mathrm{\scriptscriptstyle ICCO}}
\def\rosco{r_\mathrm{\scriptscriptstyle OSCO}}
\def\risco{r_\mathrm{\scriptscriptstyle ISCO}}

\clearpage
\section{Introduction}
\label{S:intro}

There are many physically interesting situations where a positive cosmological constant, 
\emph{no matter how small}, introduces \emph{qualitatively new} effects into general relativity,  astrophysics, and  cosmology~\cite{Ashtekar:2017}. 
One place where \emph{qualitatively new} physics arises is in the existence of \emph{outermost stable circular orbits} (OSCOs). 

Some previous work on these OSCOs has been performed both in quite abstract settings~\cite{Howes:1979, Stuchlik:1999, Lake:2019}, and within the more limited frameworks of accretion disks~\cite{Rezzolla:2003, Stuchlik:2008}, tori~\cite{Stuchlik:2009}, galaxies~\cite{Stuchlik:2011, Sarkar:2014}, ring systems~\cite{Bannikova:2018}, and axisymmetric spacetimes~\cite{Beheshti:2015}. Further afield, related calculations have also been reported 
in modified gravity~\cite{Perez:2013}, inhomogeneous FLRW cosmologies~\cite{Perez:2019},  and higher dimensions~\cite{Igata:2010,Igata:2013,Igata:2014}, but we feel there is still more to be said in this regard.
In this article we shall emphasize simple calculations, robust estimates of the relevant distances scales, and the broad astrophysical relevance of these OSCOs. 

In particular we shall compare and contrast 
the distance scale set by these OSCOs in the quasi-Newtonian, Paczy\'nski--Wiita~\cite{Paczynski:1979,Abramowicz:2009}, and general relativistic analyses.
We shall also compare these OSCOs  
with the 
distance scale set by the Jeans instability. While the underlying physics is very different (cosmological constant versus thermal pressure gradients) we shall nevertheless see that there are interesting connections between these ideas.

\section{Background}
Schwarzschild spacetime is well known to have an ISCO (innermost stable circular orbit) for massive particles at $r=6m$, and a two-sided unstable photon orbit at $r=3m$. 
What happens if we now add a (positive) cosmological constant, and consider Schwarzschild-de~Sitter (Kottler) spacetime?

One can recast  the original Kottler metric~\cite{Kottler:1918} in the form:
\begin{equation}
ds^2 = - \left(1-{2m\over r}-{r^2\over\ell^2}\right) dt^2 + {dr^2\over1-{2m\over r}-{r^2\over\ell^2}}+ r^2(d\theta^2+\sin^2\theta \, d\phi^2).
\end{equation}
As usual $m = G_N \, m_\mathrm{physical}/c^2$ is the central mass expressed as a distance, whereas $\Lambda = {3\over\ell^2}$ re-expresses the (positive) cosmological constant as an equivalent distance scale.
Locating the (massive particle) ISCOs and OSCOs exactly requires solving a quartic; but we shall see that extremely good estimates for the locations of the ISCOs/OSCOs  are not too difficult to achieve. 

We shall first present a quasi-Newtonian analysis to set the scale of the effect, then very briefly consider Schwarzschild spacetime, before focussing attention on Schwarzschild-de~Sitter (Kottler) spacetime. 
Note that by using Schwarzschild-de~Sitter spacetime we are implicitly assuming that orbital periods are short compared to the Hubble time $1/H_0$; a more refined analysis might try to analyze a version of Schwarzschild-de~Sitter spacetime embedded into a FLRW cosmology.

We shall see that to very high accuracy $r_\mathrm{\scriptscriptstyle OSCO} \approx \sqrt[3]{m \ell^2/4}$,
and we shall then estimate the size of these OSCOs for a number of astrophysically relevant objects. 

One particular reason this is astrophysically interesting is that for galactic masses $m\sim {1\over 20} \hbox{ parsec}$, and reasonable estimates for the cosmological constant $\ell \sim 5  \hbox{ Gigaparsec}$, the radius of the OSCOs are of order $r_\mathrm{\scriptscriptstyle OSCO} \approx \sqrt[3]{m \ell^2/4} \sim {7\over10} \hbox{ Megaparsec}$, comparable to the distance to the Andromeda galaxy, and so potentially small enough to affect galaxy clustering. 

\section{OSCOs in quasi-Newtonian physics}

We first start with a simple quasi-Newtonian argument for why an OSCO might be interesting once one has a positive cosmological constant.
Consider the potential
\begin{equation}
\Phi(r)  =- {m\over r}- {r^2\over 2\ell^2}.
\end{equation}
The $-{r^2\over2\ell^2}$ term in the potential above is a quasi-Newtonian approximation to the effect of a positive cosmological constant.
Then
\begin{equation}
\Phi'(r)  = {m\over r^2}- {r\over \ell^2} = { m\ell^2-r^3\over r^2\ell^2}.
\end{equation}
Note that $\Phi'\to0$ and changes sign at the critical radius $\rocco=\sqrt[3]{m\ell^2}$, which we shall soon see corresponds to an \emph{outermost conceivable circular orbit} (OCCO).  Furthermore
\begin{equation}
\Phi''(r)  = -{2m\over r^3}- {1\over \ell^2} =- { 2m\ell^2+ r^3\over r^3\ell^2},
\end{equation}
and 
\begin{equation}
\Phi''\left(\sqrt[3]{m\ell^2}\right)  = - {3 m\ell^2\over(m\ell^2)\ell^2} =- {3\over\ell^2}.
\end{equation}
We can construct a quasi-Newtonian argument for the existence of OSCOs by using the two conserved quantities 
\begin{equation}
E = {v^2\over2} + \Phi = {\dot r^2 + r^2 \dot\theta^2\over2}+\Phi; \qquad  \hbox{and}    \qquad L = r^2 \dot\theta;
\end{equation}
to write
\begin{equation}
{\dot r^2\over2} = E -\Phi -{L^2\over 2 r^2}.
\end{equation}
Thence, the Newtonian effective potential is
\begin{equation}
V(r) = \Phi + {L^2\over2r^2}=  -{m\over r}- {r^2\over 2\ell^2} + {L^2\over 2 r^2}.
\end{equation}
Furthermore
\begin{equation}
V'(r) = +{m\over r^2} - {r\over \ell^2} -{L^2\over r^3},
\end{equation}
and
\begin{equation}
V''(r) = -2{m\over r^3} - {1\over \ell^2} +{3L^2\over r^4}.
\end{equation}
Circular orbits correspond to $\dot r=0$ and $\ddot r=0$. So one must solve $V(r)=E$ to find $E(r,m,L,\ell)$, and solve $V'(r)=0$ to find $L(r,m,\ell)$. We have
\begin{equation}
L(r)^2 = {(m \ell^2 - r^3)r\over\ell^2}.
\end{equation}
That is,  circular orbits can exist only for $r\in(0, \rocco)$, with $\rocco=\sqrt[3]{m\ell^2}$; beyond this point the angular momentum required to support the circular orbit becomes imaginary.
This specifies the OCCO --- the \emph{outermost conceivable circular orbit.}
Substituting $L^2$ back into $V''(r)$ we have
\begin{equation}
V''(r) ={m\ell^2-4r^3\over \ell^2 r^3}.
\end{equation}
This can now be used to test the stability of these circular orbits for $r\in(0, \rocco)$. 
By solving for $V''(r)=0$ we can identify the (quasi-Newtonian) OSCO as
\begin{equation}
r_\mathrm{\scriptscriptstyle OSCO} = \sqrt[3]{m\ell^2/4}= 2^{-2/3}\; \sqrt[3]{m\ell^2} = 2^{-2/3}\; \rocco \approx 0.62996 \; \rocco.
\end{equation}
Note that at the OSCO and OCCO
\begin{equation}
V''(\sqrt[3]{m\ell^2/4}) = 0;
\qquad
\qquad
V''(\sqrt[3]{m\ell^2}) = - {3\over \ell^2}.
\end{equation}
While the OCCO is certainly unstable, the instability timescale is extremely long
\begin{equation}
\tau = {1\over\sqrt{|V''(\rocco)|}} = {\ell\over\sqrt{3}} \sim 9 \times 10^9 \hbox{ years}. 
\end{equation}
A full general relativity calculation will modify some of the details, but many qualitative features of this quasi-Newtonian analysis will survive. Note that there is no ISCO or ICCO for quasi-Newtonian gravity; that aspect of the quasi-Newtonian analysis will certainly change.

\section{ISCOs and OSCOs in Schwarzschild spacetime}

Let us very briefly discuss Schwarzschild spacetime in the usual coordinates. 
Take an affine parameterization (either timelike or null) for which
\begin{eqnarray}
g_{ab} {dx^a\over d\lambda} {d x^b\over d\lambda} 
&=& -\left(1-{2m\over r}\right) \left(dt \over d\lambda\right)^2 
+  \left\{{1\over1-{2m\over r}}  \left(dr \over d\lambda\right)^2 + r^2\left(d\phi \over d\lambda\right)^2 \right\} 
\nonumber\\ 
&=& \epsilon \in\{-1,0\}.
\end{eqnarray}
Killing symmetries imply two conserved quantities (energy and angular momentum)
\begin{equation}
\left(1-{2m\over r}\right)\left(dt \over d\lambda\right)=E; \qquad  
r^2 \left(d\phi \over d\lambda\right)=L.
\end{equation}
Thence
\begin{equation}
{1\over1-{2m\over r}}  \left\{-E^2 + \left(dr \over d\lambda\right)^2 \right\} + {L^2\over r^2}
=\epsilon.
\end{equation}
That is
\begin{equation}
\left(dr \over d\lambda\right)^2 =  E^2 + \left(1-{2m\over r}\right) \left\{ \epsilon -{L^2\over r^2}\right\}.
\end{equation}
This defines the ``effective potential''
\begin{equation}
V_\epsilon(r) = -{1\over2} \left(1-{2m\over r}\right)\left\{\epsilon - {L^2\over r^2}\right\}.
\end{equation}
\begin{itemize}
\item 
For $\epsilon=0$ (massless particles such as photons), the effective potential is
\begin{equation}
V_0(r) = {1\over2}\left(1-{2m\over r}\right)\,{ L^2\over r^2}.
\end{equation}
Note
\begin{equation}
V_0'(r) =  {L^2(3m-r)\over r^4},
\end{equation}
and
\begin{equation}
V_0''(r) = {3L^2(r-4m)\over r^5}.
\end{equation}
This effective potential has a single peak at $r=3m$. 
At the peak
\begin{equation}
V_0''(3m) =  -{L^2\over 81 m^4} < 0.
\end{equation}
Thus there is an unstable photon sphere at $r=3m$. 
This photon orbit is unstable in both directions. 

\item
For $\epsilon=-1$ (massive particles such as atoms, asteroids, planets, or even larger objects), the  effective potential is
\begin{equation}
V_1(r) = {1\over2}\left(1-{2m\over r}\right)\left\{1 + {L^2\over r^2}\right\}.
\end{equation}
Furthermore
\begin{equation}
V_1'(r) =  {(3mL^2-rL^2+mr^2)\over r^4},
\end{equation}
and
\begin{equation}
V_1''(r) = -{12mL^2-3rL^2+2mr^2\over r^5}.
\end{equation}
Let us solve $V'(r)=0$ for $L$:
\begin{equation}
L(r)^2 = {mr^2\over r-3m}.
\end{equation}
This has viable solutions ($L$ real and finite) for $r\in(3m,\infty)$. So there is an \emph{innermost conceivable circular orbit} (ICCO) at $r=3m$, (the photon sphere), corresponding to infinite angular momentum for the massive particle. 
Then at this value for the angular momentum one has
\begin{equation}
V_1''(r) \to {m\over r^3} {(r-6m)\over(r-3m)} .
\end{equation}
Note $V_1''(r)\to0$ and changes sign at $r=6m$, this is the standard Schwarzschild ISCO for massive particles. 

\end{itemize}

\section{The Paczy\'nski--Wiita potential}
The Paczy\'nski--Wiita potential~\cite{Paczynski:1979,Abramowicz:2009} 
\begin{equation}
\Phi(r)  =- {m\over r-2m },
\end{equation}
is a completely \emph{ad hoc} but surprisingly effective way of mimicking  ISCO/ICCO behaviour in a quasi-Newtonian context. 
Adding angular momentum we have the effective potential
\begin{equation}
V(r) = \Phi + {L^2\over2r^2} = - {m\over r-2m } + {L^2\over2r^2},
\end{equation}
with
\begin{equation}
V'(r) =  {m\over (r-2m)^2} - {L^2\over r^3},
\end{equation}
and
\begin{equation}
V''(r) =  -{2m\over (r-2m)^3} + {3L^2\over r^4}.
\end{equation}
Solving $V'(r)=0$ for $L(r)$ we have
\begin{equation}
L(r)^2 = {mr^3 \over (r-2m)^2}.
\end{equation}
We thus have a real and finite angular momentum for the two branches $r\in(2m,\infty)$ and $r\in(0,2m)$. 
We should not trust the Paczy\'nski--Wiita analysis for $r<2m$ since that would be inside the Schwarzschild radius, and it is a well-known limitation of the Paczy\'nski--Wiita analysis that it should not be fully trusted once orbital speeds approach or exceed the speed of light.

If instead we focus on the $r\in(2m,\infty)$ branch, then there is an ICCO at $\ricco=2m$ where the angular momentum diverges. (The ICCO is not at $3m$ where it would be for Schwarzschild). 

The fact that the ICCO is not exactly at the right place is a side-effect of the fact that the 
Paczy\'nski--Wiita analysis is entirely non-relativistic.  Nevertheless it is impressive to see just how good a job the Paczy\'nski--Wiita potential does. 

Inserting $L(r)$ back into $V''(r)$ we see
\begin{equation}
V''(r) \to  {m(r-6m) \over r (r-2m)^3}. 
\end{equation}
So we can identify the ISCO for the Paczy\'nski--Wiita potential as $\risco = 6m$, \emph{exactly} as for the Schwarzschild geometry.

\section{ISCOs and OSCOs for Kottler spacetime}
\def\L{{\mathcal{L}}}

With both the quasi-Newtonian and Schwarzschild discussions now in hand, the analysis for Kottler spacetime is in principle straightforward --- the only tricky issue is to find suitable approximate roots for certain quartic equations. 
To determine the ISCOs and OSCOs, consider the affinely parameterized tangent vector to the worldline of a massive or massless particle 
\begin{eqnarray}
g_{ab} {dx^a\over d\lambda} {d x^b\over d\lambda} 
&=& -\left(1-{2m\over r}-{r^2\over\ell^2}\right) \left(dt \over d\lambda\right)^2 
+  {1\over1-{2m\over r}-{r^2\over\ell^2} }\left(dr \over d\lambda\right)^2 + r^2\left(d\phi \over d\lambda\right)^2 
\nonumber
\\
&=& \epsilon \in\{-1,0\}.\quad
\end{eqnarray}

The Killing symmetries again imply the existence of two conserved quantities, (the energy and angular momentum),
\begin{equation}
\left(1-{2m\over r}-{r^2\over\ell^2}\right)\left(dt \over d\lambda\right)=E; \qquad  \hbox{and} \qquad
 r^2 \left(d\phi \over d\lambda\right)=L. 
\end{equation}
Thence
\begin{equation}
{1\over 1-{2m\over r}-{r^2\over\ell^2}} \left\{-E^2 + \left(dr \over d\lambda\right)^2 \right\} +  {L^2\over r^2}
=\epsilon.
\end{equation}
That is
\begin{equation}
\left(dr \over d\lambda\right)^2 =  E^2 + \left(1-{2m\over r}-{r^2\over\ell^2}\right)\left\{ \epsilon - {L^2\over r^2}\right\}.
\end{equation}
This defines the ``effective potential''
\begin{equation}
V_\epsilon(r) = -{1\over2} \left(1-{2m\over r}-{r^2\over\ell^2}\right) \left\{\epsilon - {L^2\over r^2}\right\}.
\end{equation}

\subsection{Unstable circular orbit for massless particles}
For $\epsilon=0$ (massless particles such as photons), the effective potential is
\begin{equation}
V_0(r) = {1\over2} {(1-2m/r-r^2/\ell^2) L^2\over r^2} = {(1-2m/r) L^2\over 2 r^2} -{L^2\over2\ell^2}.
\end{equation}
There is a single peak at $r=3m$, corresponding to 
\begin{equation}
V_{0,max} = {L^2\over54 m^2}- {L^2\over2\ell^2}, \qquad\hbox{and}\qquad V''_{0,max} = -{L^2\over81 m^4}<0.
\end{equation}
Thus there is an unstable photon sphere at $r=3m$. (This is exactly what we found for Schwarzschild.)
For massless particles there is a single circular orbit, which is unstable in both directions.
\subsection{ISCOs/OSCOs for massive particles}
For $\epsilon=-1$ (massive particles such as atoms, asteroids, or planets), we see that the  effective potential is
\begin{equation}
V_1(r) =  {1\over2} \left(1-{2m\over r}-{r^2\over\ell^2}\right) \left\{1 + {L^2\over r^2}\right\} 
= {1\over2} \left(1-{2m\over r}\right) \left\{1 + {L^2\over r^2}\right\} - {r^2\over2\ell^2} -{L^2\over2\ell^2}.
\end{equation}
We note
\begin{equation}
V_1'(r) =  {3mL^2-rL^2+mr^2-r^5/\ell^2\over r^4} =  {3mL^2-rL^2+mr^2\over r^4} -{r\over\ell^2},
\end{equation}
and
\begin{equation}
V_1''(r) =   -{12mL^2-3rL^2+2mr^2+r^5/\ell^2\over r^5} = -{12mL^2-3rL^2+2mr^2\over r^5} -{1\over\ell^2}.
\end{equation}
Finding circular orbits, ($\dot r=0$ and $\ddot r=0$), as a function of $(E,L,m,\ell)$ involves solving a quintic $V'(r)=0$ for $r(L,m,\ell)$, which is not analytically feasible, so we rearrange the calculation as follows.
Finding $L$ by solving $V'(r)=0$ as a function of $(r,m,\ell)$ is much easier:
\begin{equation}
L^2 ={r^2(m\ell^2-r^3)\over\ell^2(r-3m)}.
\end{equation}
The angular momentum is real and finite for $r\in\left(\ricco, \rocco\right]$, from the minimum conceivable circular orbit (which coincides with the unstable photon orbit at $r=3m$),  out to a maximal conceivable orbit at $\rocco=\sqrt[3]{m\ell^2}$. 
Substituting $L^2$ back into $V_0''(r)$ we see
\begin{equation}
V_1''(r) = {m(r-6m)\over r^3(r-3m)} -{(4r-15m)\over(r-3m)\ell^2} 
= {-4r^4+15 mr^3+m\ell^2 r -6m^2\ell^2\over (r-3m) r^3 \ell^2} .
\end{equation}
Finding the ISCO and OSCO by solving $V_1''(r)=0$ exactly requires solving a quartic; this is in principle do-able but in practice not particularly useful.
Noting that 
\begin{equation}
V_1''(r=6m) = -{3\over \ell^2} <0, \qquad \hbox{and} \qquad
V_1''(\rocco) = -{3\over \ell^2} <0,
\end{equation}
we see that the orbits at $6m$ and $\rocco$ are unstable, so that the ISCO lies somewhere \emph{above} $6m$ and the OSCO lies somewhere \emph{below} $\rocco$. 
The instability timescale is 
\begin{equation}
\tau = {1\over\sqrt{|V_1''|}} \sim {\ell\over\sqrt{3}} \sim 9\times10^9 \hbox{ years}, 
\end{equation}
so for most practical purposes we can get away with using $6m$ for the ISCO and $\rocco$ for the OSCO.
Let us now find approximate but robust esimates for the location of the ISCO and OSCO using semi-analytic methods.

\subsubsection{Approximate location of the ISCO}
To estimate the location of the ISCO we simply employ one iteration of the Newton--Raphson method. Knowing that the effect of the cosmological constant on gravitational dynamics is negligible near the centralised mass, we may assume an initial guess for the location of the ISCO is given by the Schwarzschild value $\risco \approx r_{0}=6m$. The first  Newton--Raphson iteration yields:
\begin{eqnarray}
r_{1} &=& r_{0}-\frac{V_{1}^{''}\left(r_{0}\right)}{V_{1}^{'''}\left(r_{0}\right)} 
= 6m + \frac{1944\,m^{3}}{\ell^2+432\,m^2} 
= 6m\left[1+\frac{324\,m^2}{\ell^2+432\,m^2}\right].
\end{eqnarray}
Note that this is very close to $6m$, but (as expected) slightly further away from the central mass.
Furthermore, we note
\begin{equation}
V_1''(r_1) = -{13608\,m^2\over \ell^4}+O\left(m^4\over\ell^6\right) = 4536\; V_1''(r_0) {m^2\over \ell^2} +O\left(m^4\over\ell^6\right).
\end{equation}
That is, $r_1$ is certainly a much better estimate for the location of the ISCO defined by $V_1''(\risco)=0$ than the initial guess $r_0=6m$. 
Accordingly we estimate the approximate location for the ISCO as
\begin{equation}
    \risco \approx 6m\left[1+\frac{324m^2}{\ell^2+432m^2}\right] 
    = 6m\left[1+\frac{324m^2}{\ell^2} + O\left( {m^4\over\ell^4}\right) \right].
\end{equation}
In most astrophysically interesting situations this correction to the naive $\risco\approx 6m$ is so small as to be quite negligible. 

\subsubsection{Approximate location of the OSCO}

In contrast, to determine the approximate location of our OSCO, the Newton--Raphson technique proves insufficient. Instead  we adopt the following method: Let us define $r = x\cdot \rocco = x\cdot \sqrt[3]{m\ell^2}$ and then write
\begin{equation}
V_{1}^{''}\left(x, \ell^2\right) = \frac{-4\left(m^2\ell\right)^{\frac{2}{3}}x^{4}+15m^2x^3+\left(m^2\ell\right)^{\frac{2}{3}}x-6m^2}{r^3\left(r-3m\right)} .
\end{equation}
Now  perform a series expansion around $\ell^2=+\infty$, corresponding to $\Lambda=0$. We find
\begin{equation}
V_{1}^{''}\left(x, \ell^2\right) = \frac{-8x^{4}+2x}{x^{4}\ell^2} +\frac{1}{\ell^{2}}\cdot O\left[\left({m}/{\ell}\right)^{{2}/{3}}\right] \ .
\end{equation}
The presence of fractional powers means that this not a Taylor series expansion. It is instead a ``generalized power series''~\cite{Cattoen:2005}, or ``generalized Frobenius series''~\cite{Visser:2002}, also called a Puiseaux series~\cite{Puiseux,Cattoen:2006,Cattoen:2007}.

We can clearly see that the $O\left[\left({m}/{\ell}\right)^{{2}/{3}}\right]$ term is subdominant, and as such may simply focus on the values of $x$ which make the dominant term vanish: $2x(4x^3-1)=0$. This corresponds to $x\in\{0, 2^{-2/3}\}$. 
We may immediately discount the unphysical root at $x=0$, and take $x= 2^{-2/3}$.
Note
\begin{equation}
V_{1}^{''}\left(x= 2^{-2/3}, \ell^2\right) =
-{18m\over \ell^2 \left(\sqrt[3]{m\ell^2/4}-3m\right)} = -{3\over \ell^2}\; {6m\over\sqrt[3]{m\ell^2/4}-3m}.
\end{equation}
Thus $V_{1}^{''}\left(x= 2^{-2/3}, \ell^2\right)$, while nonzero, is certainly extremely small compared to $-3/\ell^2$, and the corresponding instability timescale is extremely large, much larger than the age of the universe. That is, for all practical purposes an adequate approximation to the location of the OSCO is to take
\begin{equation}
\rosco \approx 2^{-2/3} \, \rocco = \sqrt[3]{m\ell^2/4}.
\end{equation}
Note that the general relativity calculation (approximately) matches the quasi-Newtonian calculation. 

\section{Paczy\'nski--Wiita potential with cosmological constant}

For completeness, let us now add a cosmological constant to the  Paczy\'nski--Wiita potential~\cite{Paczynski:1979,Abramowicz:2009}. We set
\begin{equation}
\Phi(r)  =- {m\over r-2m } - {r^2\over 2\ell^2}.
\end{equation}
This is completely \emph{ad hoc}, but as we shall soon see, this is a surprisingly effective way of mimicking  both ISCO/ICCO and OSCO/OCCO behaviour in a quasi-Newtonian context. 

Now adding angular momentum we have the effective potential
\begin{equation}
V(r) = \Phi + {L^2\over2r^2} = - {m\over r-2m } - {r^2\over 2\ell^2} + {L^2\over2r^2},
\end{equation}
with
\begin{equation}
V'(r) =  {m\over (r-2m)^2} -{r\over \ell^2} - {L^2\over r^3},
\end{equation}
and
\begin{equation}
V''(r) =  -{2m\over (r-2m)^3} - {1\over\ell^2} + {3L^2\over r^4}.
\end{equation}
Solving $V'(r)=0$ for $L(r)$ we have
\begin{equation}
L(r)^2 = {mr^3 \over (r-2m)^2}-{r^4\over\ell^2}.
\end{equation}
As per the previous discussion, (which applied only in the absence of a cosmological constant), from this we deduce the presence of an ICCO at $\ricco=2m$, where $L(r)\to\infty$.
There is now also an OCCO near $\rocco \approx \sqrt[3]{m\ell^2}$, where $L(r)$ becomes imaginary.  

Inserting $L(r)$ back into $V''(r)$ we see
\begin{equation}
V''(r) \to  {m(r-6m) \over r (r-2m)^3} - {4\over\ell^2}. 
\end{equation} 
The stability region for these circular orbits is determined by finding the (approximate) roots of $V''(r)=0$.

Accordingly the ISCO is near $\risco\approx 6m$ and the OSCO near $\rosco \approx \sqrt[3]{m\ell^2/4}$. This is all qualitatively and quantitatively very similar to what we saw happened for the Kottler spacetime. We could try to further refine the locations of these roots (by Newton--Raphson or other means) but given what we have already seen happens  for the Kottler spacetime, such further refinements seem unnecessary.

\section{Some astrophysical estimates}

We now provide some OSCO estimates for astrophysically interesting situations; this requires using observational estimates for both the cosmological constant and for the relevant astrophysical masses.

First, we set $\ell= 5$ Gigaparsec, corresponding to $\Lambda = {3\over\ell^2} = 1.2 \times 10^{-19} \hbox{ parsec}^{-2}$. 
Note that while $\Omega_\Lambda= 0.692 \pm 0.012$ is known at the 1\% level, the roughly 10\% discrepant estimates of the Hubble parameter $H_0$, and consequently 20\% discrepant estimates of the Hubble density $\rho_H = 3H_0^2/(8\pi G_N)$, imply a 20\% uncertainty in the cosmological constant $\Lambda$, and a 10\% uncertainty in the cosmological distance scale $\ell$. For this reason there is currently no point in asserting more precision than  $\ell= 5$ Gigaparsec.

In Table~1 we present a few typical masses (in parsecs), calculate the corresponding $\rocco$, and comment on the astrophysical relevance of the resulting distance scale. Note that many astrophysical masses are uncertain up to factors of 2, some are uncertain up to factors of 10. Accordingly we typically work to only 1 or at most 2 significant figures.
A remarkable aspect of the estimates reported in Table~1 is that the OSCOs are not cosmologically large; indeed some of them are suspiciously close to distance scales relevant to the observed hierarchical behaviour of structure formation in the universe. 

More specifically,  we emphasise:
\begin{itemize}
\itemsep0pt
\item 
For atomic masses the OSCO is of order 20 centimetres. 
\item 
For planetary masses the OSCO is of order a parsec. 
\item
For galactic masses the OSCOs are of order the inter-galactic spacing,
\item  
For galaxy cluster masses the OSCOs are of order the size of the cluster. 
\end{itemize}

\clearpage. 
\begin{table}[!ht]
\begin{center}
\caption{OSCOs as a function of mass ($\ell = 5 \hbox{ Gpc}$).}
\end{center}
\hspace{-35pt}
\begin{tabular}{||c|c|c||c||}
\hline
\hline
\vphantom{\bigg|} Object & $m$ (in parsec)  & $\rosco  = \sqrt[3]{m\ell^2/4}$ (in parsec) & Astrophysical Relevance? \\
\hline
\hline
Hydrogen atom &  $4.2 \times 10^{-71}$ &  $6.5\times 10^{-18}$ \quad (20 cm) & Dust clouds.\\
Earth &		        $1.5 \times 10^{-19}$ & $1$ & Rogue planets.\\
Sun &		 	$5 \times 10^{-14}$ &  $70$ &  Scale height of galactic disk.\\
Stellar association & $5 \times 10^{-13}$ & $150$ & Size of association.\\
Open stellar cluster & $5 \times 10^{-12}$ & $300$ & Open cluster spacing.\\
Globular cluster &         $5 \times 10^{-9}$ &  $3 \times 10^3$ & Globular cluster spacing.\\
Saggitarius A$^*$ 	& $2\times 10^{-7}$ & $10^4$ & Size of galaxy.\\
Dwarf galaxies &  $5 \times 10^{-5}$ & $7 \times 10^4$ & Inter-dwarf spacing. \\
Spiral galaxies    & $5 \times 10^{-2}$ & $7 \times 10^5$ & Inter-galactic spacing.\\
Galaxy clusters       & $50$ & $7 \times 10^6$ & Size of galaxy cluster.\\
Observable universe & $2.5 \times 10^9$ & $2.5 \times 10^9$ & Observable universe.\\
\hline
\hline
\end{tabular}
\end{table}

Specifically, for atomic Hydrogen dominated dust clouds the fact that $\rocco\sim 20\; \hbox{cm}$ implies instability to cosmological-constant induced shredding once the number density drops below $n_\mathrm{critical} \sim {60} \; \hbox{m}^{-3}$. 
For molecular clouds this effect scales slowly with the cube root of the average molecular mass.

Here is another way of looking at these results in a somewhat more general setting:
Let us consider an arbitrary gravitationally bound object, of total mass $m$ and size $r_*$, then certainly $r_*\leq\rosco$ and so the average density satisfies
\begin{equation}
\bar \rho = {m\over{4\pi\over 3} r_*^3} \geq {m\over{4\pi\over 3} \rosco^3} 
= {m\over{4\pi\over 3} (m\ell^2/4)} ={3\over\pi\ell^2} 
= {\Lambda\over\pi} = 8 \rho_\Lambda.
\end{equation}
That is, $\bar\rho\geq 8\rho_\Lambda$; a positive cosmological constant, no matter how small, induces an over-density gap

 That is, all gravitationally bound objects must exhibit  an average over-density considerably higher than the equivalent energy density $\rho_\Lambda$ associated with the cosmological constant. Diffuse systems are likely to be near saturating this bound, implying $r_* \lesssim \rosco$. Compact systems will individually exhibit $r_* \ll \rosco$, however \emph{collections} of compact systems are likely to be diffuse, with size $r_* \lesssim \rosco$, implying a spacing between compact objects of order $\rosco$. 

\section{Comparison with the Jeans scale}\label{S:Jeans}

We shall now briefly discuss the Jeans length as being \emph{complementary} to the OSCO/OCCO scale,  giving you somewhat \emph{different} information.
\begin{itemize}
\item Comparing clouds with the same density and same speed of sound, the Jeans length sets the \emph{minimum} scale for gravitational clustering in a gas cloud that has a well-defined speed of sound.
\item The OSCO/OCCO sets a maximum scale for gravitational clustering,
and does not care about the speed of sound.
\end{itemize}

One rather common formula (in physical units) for the Jeans length in terms of speed of sound $c_s$ and density $\rho$ is this:
\begin{equation}
\lambda_J \sim {c_s\over \sqrt{G\rho}}.
\end{equation}
This criterion comes from comparing the sound crossing time $t_s= r/c_s$ with the Newtonian gravitational free-fall time $t_\mathrm{ff} = 1/\sqrt{G\rho}$. 
Here  $c_s^2 = \partial p/\partial\rho$. 
For a system in internal thermal equilibrium $c_s^2 \sim k_B T/\bar\mu$, where $\bar \mu$ is the average molecular mass. 
Note that the Jeans criterion makes sense only if the system of interest is more-or-less homogeneous. 

When comparing different size clouds with the same $\rho$ and same $c_s$ the system is pressure dominated (and so uncollapsed) for $r \lesssim \lambda_J$, but unstable to gravitational collapse for $r\gtrsim\lambda_J$. 
Now let us rephrase this in a manner more useful for our current purposes:
If the system has total mass $m$ then using $\rho \sim m/r^3$ we can rewrite the condition for pressure domination as
\begin{equation}
r \lesssim {c_s\over \sqrt{Gm/r^3}}.
\end{equation}
Thence
\begin{equation}
r^2 G (m/r^3) \lesssim c_s^2.
\end{equation}
That is 
\begin{equation}
r\gtrsim {G m\over c_s^2}.
\end{equation}

So when re-phrased in terms of comparing different size clouds with the same speed of sound and the same total mass, pressure dominance corresponds to a diffuse system
\begin{equation}
r \gtrsim \lambda_J \sim {G m\over c_s^2},
\end{equation}
and gravitational collapse corresponds to a compact system
\begin{equation}
r \lesssim \lambda_J \sim {G m\over c_s^2},
\end{equation}
with the Jeans stability criterion switching at
\begin{equation}
r \sim \lambda_J \sim {G m\over c_s^2},
\end{equation}
The naive apparent switch in the \emph{direction} of the Jeans inequality is a subtle one, and has to do with what is being held fixed as one compares dust clouds with each other.

Therefore we have the perhaps unexpected relation
\begin{equation}
\lambda_J \sim  \hbox{(Schwarzschild radius)} \times
 {\hbox{(speed of light)}^2\over \hbox{(speed of sound)}^2}.
\end{equation}
allowing us to interpret the Jeans length as an ``acoustic analogue'' of the Schwarzschild radius~\cite{analogue1,analogue2}.

In short, when assuming the cosmological constant is zero, in order for there to be non-empty region of pressure-induced stability one must have $c_s > 0$.

We now wish to consider the effect of non-zero positive cosmological constant, and compare and contrast the Jeans scale with the OCCO,
which in physical units is
\begin{equation}
\rocco =\sqrt[3]{(Gm/c^2)\ell^2}.
\end{equation}
In theoreticians units we would set $G\to1$ and use $c_s$ to denote the dimensionless ratio of sound-speed to light-speed so that
\begin{equation}
\lambda_J \sim {m\over c_s^2}; \qquad \rocco =\sqrt[3]{m\ell^2}.
\end{equation}

Then  when comparing different size clouds of fixed mass and fixed speed of sound:
\begin{itemize}
\item 
For compact clouds $r\in(0,\lambda_J)$ one has gravitational collapse.
\item 
For medium scale clouds $r\in(\lambda_J,\rocco)$ one has stability.
\item 
For diffuse clouds $r\in(\rocco, \infty)$ one has cosmological-constant induced shredding.
\end{itemize}
For the region of stability to be non-empty one requires
\begin{equation}
\lambda_J <\rocco.
\end{equation}
That is
\begin{equation}
{m\over c_s^2} \lesssim \sqrt[3]{m\ell^2}.
\end{equation}
This implies
\begin{equation}
{m^3\over c_s^6} \lesssim {m\ell^2}.
\end{equation}
Consequently in order for there to be non-empty region of pressure-induced stability in the presence of a positive cosmological constant one must have
\begin{equation}
c_s \gtrsim \sqrt[3]{m\over \ell}; \qquad\hbox{equivalently}\qquad m \lesssim \ell\; c_s^3 .
\end{equation}

To set the scale for this effect, with very few exceptions temperatures in the universe are bounded below by that of the CMB, so $T \gtrsim T_{CMB} \approx 2.7 \hbox{ K}$. 

For molecular Hydrogen this corresponds to 
\begin{equation}
c_s \sim \sqrt{k_B T \over 2 m_H} \sim 3.4 \times 10^{-7} \to 100 \hbox{ m/s}.
\end{equation}
Then even at this lowest plausible temperature, taking $m \lesssim 4000$ solar masses is enough to guarantee a non-empty interval of pressure-domination. 

\section{Discussion and Conclusions}\label{S:Conclusions}
There are many situations in which the presence of an arbitrarily small positive cosmological constant leads to \emph{qualitatively} new phenomena~\cite{Ashtekar:2017}. 
One such situation is the presence of OSCOs (\emph{outermost stable circular orbits}) a phenomenon that occurs in the presence of a positive cosmological constant. 

(No such effect arises for negative cosmological constant; these OSCO effects are not an issue in asymptotically anti-de~Sitter space.) We have analyzed the existence and sizes of OSCOs in both quasi-Newtonian gravity and in standard general relativity (the Schwarzschild-de~Sitter [Kottler] black hole),
and have developed simple robust arguments for the existence of OCCOs and OSCOs, and simple robust estimates for the location of these OSCOs. The most interesting part of the analysis is that the OSCOs are \emph{not} cosmologically large, they are small enough to be astrophysically interesting. 

\acknowledgments{
This project was funded by the Ratchadapisek Sompoch Endowment Fund, Chulalongkorn University (Sci-Super 2014-032), by a grant for the professional development of new academic staff from the Ratchadapisek Somphot Fund at Chulalongkorn University, by the Thailand Research Fund (TRF), and by the Office of the Higher Education Commission (OHEC), Faculty of Science, Chulalongkorn University (RSA5980038). PB was additionally supported by a scholarship from the Royal Government of Thailand. TN was also additionally supported by a scholarship from the Development and Promotion of Science and Technology talent project (DPST). \\
AS acknowledges financial support via the PhD Doctoral Scholarship provided by Victoria University of Wellington.
AS is indirectly supported by the Marsden fund, administered by the Royal Society of New Zealand.
MV was supported by the Marsden Fund, via a grant administered by the Royal Society of New Zealand.
}


\end{document}